%% file: Proceeding_Outflows_astroph.tex
\documentclass{cupbook}

\input{journals}

\usepackage{epsf}
\usepackage{graphicx}
\usepackage{./natbib}

\def\eps@scaling{.98}

\def\showone#1{
  \centering
  \leavevmode
  \epsfxsize=\eps@scaling\linewidth
  \epsfbox{#1.eps}
}

\def\epsvert@scaling{.45}
\def\showonevert#1{
  \centering
  \leavevmode
  \epsfysize=\epsvert@scaling\textheight
  \epsfbox{#1.eps}
}

\def\epstwo@scaling{0.49}
\def\showtwo#1#2{
  \centering
  \leavevmode
  \epsfxsize=\epstwo@scaling\linewidth
  \epsfbox{#1.eps} \hfil
  \epsfxsize=\epstwo@scaling\linewidth
  \epsfbox{#2.eps}
}

\newcommand{\sm}[1]{\mbox{{\scriptsize #1}}}

\newcommand{\bef}{\begin{figure}}
\newcommand{\eef}{\end{figure}}

\newcommand{\cm}{\mbox{cm}}

\newcommand{\km}{\mbox{km}}

\renewcommand{\sec}{\mbox{s}}
\newcommand{\g}{\mbox{g}}

\newcommand{\sun}{\odot}
\newcommand{\Msol}{\mbox{$M_{\sun}$}}

\newcommand{\ys}{\mbox{years}}

\begin{document}

\chapter{The role of jets in the formation of planets, stars, and galaxies}

Ralph E.~Pudritz$^1$, Robi Banerjee$^2$ and Rachid Ouyed$^3$ \\
$^1$ {\small{\it Kavli Institute for Theoretical Physics, University of California, Santa Barbara, CA 93106-4030}} \\
$^2$ {\small{\it Insitut f\"ur Theoretische Astrophysik, Universit\"at
    Heidelberg, 69120 Heidelberg, Germany}} \\
$^3$ {\small{\it Department of Physics and Astronomy, University of
    Calgary, Calgary, Alberta T2N 1N4, Canada}}

\subsection{Abstract} 

Astrophysical jets are associated with the formation of young stars of
all masses, stellar and massive black holes, and perhaps even with the
formation of massive planets.  Their role in the formation of planets,
stars, and galaxies is increasingly appreciated and probably reflects
a deep connection between the accretion flows - by which stars and
black holes may be formed - and the efficiency by which magnetic
torques can remove angular momentum from such flows.  We compare the
properties and physics of jets in both non-relativistic and
relativistic systems and trace, by means of theoretical argument and
numerical simulations, the physical connections between these
different phenomena.  We discuss the properties of jets from young
stars and black holes, give some basic theoretical results that
underpin the origin of jets in these systems, and then show results of
recent simulations on jet production in collapsing star-forming cores
as well as from jets around rotating Kerr black holes.

\section{Introduction}

The goal of this book, to explore structure formation in the cosmos
and the physical linkage of astrophysical phenomena on different
physical scales, is both timely and important.  The emergence of
multi-wavelength astronomy in the late 20th century with its
unprecedented ground and space-based observatories, as well as the
arrival of powerful new computational capabilities and numerical
codes, have opened up unanticipated new vistas in understanding how
planets, stars, and galaxies form.  Galaxy formation has turned out to
be a complex problem which requires a deep understanding of star
formation on galactic scales and how it feeds back on galactic gas
dynamics.  The discovery of significant numbers of massive black holes
in galactic nuclei now suggests that most galaxies harbour million to
billion mass holes.  The formation of these monsters and the effects
that they can have on the early evolution of galaxies is still not
well understood however.  Star formation studies have made huge
inroads, but must still take a large step before we can truly
understand the origin of stellar masses and star formation rates in
molecular clouds and galaxies.  The newly discovered planetary systems
bear little relation to text-book models of how our solar system is
believed to have formed.  It is also abundantly clear that planet and
star formation are intimately coupled through the physics of
protostellar disks that are their common cradles.

Astrophysical jets play an important role in the formation and
evolution of stars, black holes, and perhaps massive planets.  They
appear to be an inescapable multi-scale phenomenon that arises during
structure formation. The enormous kinetic luminosity of many
jets (often comparable to the bolometric luminosity of the central
sources) also implies that they could have important feedback effects on
structure formation; from the scale of the giant lobes of radio
galaxies stretched out across many Mpc scales in the IGM, to jets from
stellar mass black holes, down to the stirring up of molecular gas
on sub-pc to pc scales in regions of low mass star formation.  Jets are
observed to be ubiquitous during the process of star formation and
associated with a very broad range of objects; from brown dwarfs to B
and perhaps even O stars.  It is now well established that stellar
mass black holes are associated with jets and that the properties of
such "microquasars" scale very well only with the mass of the black
hole \citep[e.g. see review,][]{Mirabel05}, and scale naturally to the limit
of massive black holes.  In this regard, galactic jets are 
being increasingly associated with accreting massive black holes
\citep[e.g.  review,][]{Blandford01}.

In most cases, there is clear evidence that accretion
disks are an essential part of the engine.  The fact that outflows and
jets are observed around nearly all astrophysical objects during the
early stages of their existence - a time where the central objects
(young stars, massive planets, or black holes) undergo significant
accretion from surrounding collapsing gas structures and/or associated
disks - argues for a deep link between accretion and outflow.
A large body of observations and theoretical models increasingly
suggests that such jets in all of these systems may be powered by the
same mechanism, namely, hydromagnetic winds driven off of magnetised
accretion disks \citep[see e.g.][for reviews on this subject]{Livio99,
Konigl00, Pudritz07}.  The basic theory for hydromagnetic disk winds
was worked out by \citet{Blandford82} in the context of models for AGN
jets, but was quickly found to be very important for understanding
jets in protostellar systems (PN83).  For accreting systems, the
gravitational binding energy that is released as gas accretes through
a disk and onto a central object is the ultimate source of the energy
for jets.  Therefore, if this energy can be efficiently tapped and
carried by the jet, then jet energies and other properties should
simply scale with the depth of the gravitational potential well
created by the central mass.  The torques that can be exerted by
outflows upon the underlying disks were shown to be much larger than
can be produced by even strongly turbulent disks \citep[][]{Pudritz83,
Pelletier92}.  Thus, jets can also carry off most of the angular
momentum underlying disks, thereby assisting in the growth of central
objects - be they stars, black holes, or planets.  (It should be noted
that large scale spiral waves in accretion disks can also efficiently
transport angular momentum through disks).  Finally, the feedback of
these powerful jets upon their surroundings can be important.  In star
forming systems, outflows may help to drive turbulence within cluster
forming regions in molecular clouds, which would help to sustain such
a region against global collapse.

It must be noted that isolated magnetised, spinning bodies such as
magnetised A stars can also drive drive outflows.  The extraordinary
Chanra X-ray observations of Crab pulsar show that it driving off a
highly collimated, relativistic jet.  The models that best describe
the physics of all of these jets demonstrate that outflows can be
driven and collimated from spinning bodies (be they disks, stars, or
compact objects).  In protostellar jets, the slow spin of stars has
been attributed either to star-disk coupling \citep{Koenigl91} or
X-winds \citep{Shu00}, but could also be explained as an accretion
powered stellar wind from the central star \citep[e.g.,][]{Matt05}.
For black hole systems, the highly relativistic component of jets may
be associated with electrodynamic processes and currents that arise
from spinning holes in a magnetised environment provided by the
surrounding disk \citep[e.g.,][]{Blandford77}.

It is impossible to do justice to the enormous body of excellent work
on the physics of jets in these different systems.  In this chapter,
we give specific examples of these basic themes from our own research
in these fields in the context of some of the basic literature.  In \S
2 we review aspects of unified models for non-relativistic, and then
relativistic jets in these systems.  We follow this in \S 3 with a
discussion of disk winds as an excellent candidate for a unified model,
analysed particularly for protostellar jets.  We follow this in \S 4
with simulations of jets during gravitational collapse and disk
formation.  We examine the role of feedback of jets in the specific
context of star formation in \S 5.  We then move on to the basics of
relativistic jets from compact objects as well as black holes (\S 6),
and then simulations (\S 7) of MHD jets in the general relativistic
limit (GRMHD).  We conclude with a comparison of the physics of 
relativistic vs non-relativistic jets.

\section{Jets in diverse systems }

\textbf{Protostellar objects }: Stars span about 4 decades in mass.
Until recently, the study of outflows was restricted to sources that
were easily detectable in millimetre surveys of molecular clouds.
These studies show that
outflows, over a vast range of stellar luminosities,
have scalable physical properties.  In particular the observed ratio
of the momentum transport rate (or thrust) carried by the CO molecular
outflow to the thrust that can be provided by the bolometric
luminosity of the central star \citep[e.g.,][]{Cabrit92}; scales as
\begin{equation}
F_{\sm{outflow}}/F_{\sm{rad}} = 250 (L_{\rm bol}/10^3 L_{\odot})^{-0.3}, 
\end{equation}
This relation has been confirmed and extended by the analysis of data
from over 390 outflows, ranging over six decades up to $10^6L_{\odot}$
in stellar luminosity~\citep{Wu04}.  This remarkable result shows that
that jets from both low and high mass systems, in spite of the
enormous differences in the radiation fields from the central stars,
are probably driven by a single, non-radiative, mechanism.

Jets are observed to have a variety of structures and time-dependent
behaviour -- from internal shocks and moving knots to systems of bow
shocks that suggest long-time episodic outbursts \citep[e.g., review
][]{Bally07}.  They show wiggles and often have cork-screw like
structure, suggesting the presence either of jet precession, the
operation of non-axisymmetric kink modes, or both.  Numerical
simulations have recently shown that in spite of the fact that jets
are predicted to have a dominant toroidal magnetic field, they are
nevertheless stable against such nonlinear modes
\citep[e.g.,][]{Ouyed03, Nakamura04, Kigure05}.  The nonlinear
saturation of the unstable modes is achieved by the natural regulation
of the jet velocity to values near the local Alfv\'en speed.  A second
point is that jets will have some poloidal field along their
"backbone" and this too prevents a jet from falling apart.
 
Two types of theories that have been proposed to explain protostellar
jets, both of which are hydromagnetic; disk winds \citep[e.g.,
review][]{Pudritz07}, or the X-wind \citep[e.g.,
review][]{Shang07}.  The latter model posits the interaction between a
spinning magnetised stellar magnetosphere and the inner edge of an
accretion disk as the origin of jets, while the former envisages jets
as arising from large parts of magnetised disks.  One of the best ways
of testing such theories is by measuring jet rotation - which is a
measure of how much angular momentum that a jet can extract from its
source.  Significant observational progress has been made on this
problem lately with the discovery of the rotation of jets from T-Tauri
stars.  The angular momentum that is observed to be carried by these
rotating flows (e.g.\ DG Tau) is a considerable fraction of the excess
disk angular momentum -- from 60--100\% \citep[e.g.,][]{Bacciotti04},
which is consistent with the high extraction efficiency that is
predicted by the theoretical models.  Another important result is that
there is too much angular momentum in the observed jet to be accounted
for the spinning star or very innermost region of the disk as
predicted by the X-wind model.  The result is well explained by the
disk wind model which predicts that the angular momentum derives from
large reaches of the disk, typically from a region as large as 0.5 AU
\citep{Anderson03}.

\textbf{Jovian planets }:  The striking similarity of the system of 
Galilean moons around Jupiter, with the sequence of planets in 
our solar system, has long suggested that these moons may have formed
through a subdisk around Jupiter \citep{Mohanty07}.
Recent hydrodynamical simulations of circumstellar accretion disks        
containing and building up an orbiting protoplanetary core have
numerically proven the existence of a circum-planetary sub-disk in        
almost Keplerian rotation close to the planet \citep{Kley01}.
The accretion rate of these sub-disks is about                            
$\dot{M}_{\rm cp} = 6\times 10^{-5}\,M_{Jup}\,{\rm yr}^{-1}$
and is confirmed by many independent simulations.                         
With that, the circum-planetary disk temperature may reach values         
up to $2000\,$K indicating a sufficient degree of ionisation for          
matter-field coupling and would also allow for strong equipartition       
field strength \citep{Fendt03}.                                       
It should be possible, therefore, for lower luminosity jets to be    
launched from the disks
around Jovian mass planets. 

The possibility of a planetary scale MHD outflow, similar to 
the larger scale YSO disk winds, is indeed quite likely because:
(i) the numerically established existence of circum-planetary disks
is a natural feature of the formation of massive planets;
and (ii) the feasibility of a large-scale magnetic field in the               
  protoplanetary environment                                                
\citep{Quillen98, Fendt03}.    
One may show, moreover, that 
the outflow velocity is of the order of the escape speed for the  
 protoplanet, at about                                                     
$60\,{\rm km\,s}^{-1}$ \citep{Fendt03, Machida06}.

\textbf{Black holes }: Relativistic jets have been observed or
postulated in various astrophysical objects, including active galactic
nuclei (AGNs) \citep[e.g.,][]{Urry95, Ferrari98},
microquasars in our galaxy \citep[e.g.,][]{Mirabel98}, and
gamma-ray bursts (GRBs) \citep[e.g.,][]{Piran05}.  Table
\ref{tab:ultra-jets} summarises the features or relativistic jets in
different large scale sources and demonstrate the wide range in power
and Lorentz factors achieved by these systems.

\begin{table}[htdp]
\caption{The sources and features of large scales/relativistic jets}
\begin{center}
\begin{tabular}{|c|c|c|}\hline
 Source      &      $L$           &   $\Gamma$       \\\hline
 AGNs/Quasars      &      $10^{43}$-$10^{48}$ erg s$^{-1}$           &   $\sim 10$       \\\hline
 $\mu$-quasars      &      $10^{38}$-$10^{40}$ erg s$^{-1}$          &   1-10       \\\hline
 GRBs      &      $10^{51}$-$10^{52}$ erg s$^{-1}$           &   $10^2$-$10^4$       \\\hline
\end{tabular}
\end{center}
\label{tab:ultra-jets}
\end{table}%
 
In the commonly accepted standard model of large scale/relativistic
jets \citep{Begelman84}, flow velocities as large as 99\% of the speed
of light (in some cases even beyond) are required to explain the
apparent superluminal motion observed in many of these sources (see
Table \ref{tab:ultra-jets}).  Later, considerations of stationary MHD
flows have revealed that relativistic jets must be strongly magnetised
\citep{Michel69, Camenzind86, Li92, Fendt96}.  In that case, the
available magnetic energy can be transfered over a small amount of
mass with high kinetic energy.

Here too, the most promising mechanisms for producing the relativistic
jets involve magnetohydrodynamic centrifugal acceleration and/or
magnetic pressure driven acceleration from an accretion disk around
the compact objects \citep[e.g.,][]{Blandford82}, or involve the
extraction of rotating energy from a rotating black hole
\citep{Penrose69, Blandford77}.  These models have been applied to
explain jets features in the galactic
micro-quasars GRS 1915+105 \citep{Mirabel94} and GRO J1655-40
\citep{Tingay95}, or a rotating super massive black hole in an
active galactic nucleus, which is fed by interstellar gas and gas from
tidally disrupted stars. In general these studies require solving
special relativistic MHD (SRMHD) or general relativistic MHD (GRMHD)
equations and often require sophisticated numerical codes.  We
describe the basic theory behind relativistic MHD jets and summarise
recent GRMHD simulations of jets emanating from the vicinity of
accreting black holes (see \S 6 and 7) after we examine the general
theory of outflows from magnetised spinning disks and objects..

\section{Theory of disk winds} 

Given the highly nonlinear behaviour of the force balance equation for
jets (the so-called Grad-Shafranov equation), theoretical work has
focused on tractable and idealised time-independent, and axisymmetric
or self-similar models (e.g., BP82) of various kinds. We briefly
summarise the theory, below \citep[see details in][]{Pudritz07,
Pudritz04}.

\subsubsection{Conservation Laws and Jet Kinematics} 

Conservation laws play a significant role in understanding
astrophysical jets.  This is because whatever the details,
conservation laws strongly constrain the flux of mass, angular
momentum, and energy.  What cannot be constrained by these laws will
depend on the general physics of the disks such as on how matter is
loaded onto field lines.

Jet dynamics can be described by the time-dependent, equations of
ideal MHD.  The evolution of a magnetised, rotating system that is
threaded by a large-scale field ${\bf B}$ involves (i) the continuity
equation for a conducting gas of density $\rho$ moving at velocity
${\bf v} $ (which includes turbulence); (ii) the equation of motion
for the gas which undergoes pressure ($p$), gravitational (with
potential $\Phi$), and Lorentz forces; (iii) the induction equation
for the evolution of the magnetic field in the moving gas where the
current density is ${\bf j} = (c / 4 \pi) {\bf \nabla \times B}$; (iv)
the energy equation, where $e$ is the internal energy per unit mass;
and, (v) the absence of magnetic monopoles. These are written as:
\begin{eqnarray}
{\partial\rho\over\partial t}+\nabla .(\rho{\bf v}) &=&0 \qquad  \\
\rho \left({\partial{\bf v}\over \partial t}+({\bf v.\nabla}){\bf v}\right)
+\nabla p +\rho {\bf \nabla}\Phi -
{{\bf j}\times {\bf B}\over c}&=&0 \qquad \\
{\partial {\bf B}\over\partial t}-\nabla\times ({\bf v}\times {\bf B})&=&0 \qquad  \\
\rho\left({\partial{ e}\over \partial t}+({\bf v.\nabla)}e\right)
+ p({\bf \nabla .v})&=&0 \qquad \\
\nabla .{\bf B}&=&0 \qquad
\end{eqnarray}

Progress can  be made by restricting the analysis to
stationary, as well as  2D (axisymmetric) flows, from 
which the conservation laws follow. 
It is useful to decompose vector quantities into poloidal
and toroidal components (e.g.\ magnetic field 
$  {\bf B =  B_p} +  B_{\phi}
{\bf \hat e_{\phi}} $).  In axisymmetric conditions, the 
poloidal field ${\bf B_p}$ can be derived from a single scalar potential
$a(r,z)$ whose individual values, $a=\mbox{const}$, define the 
surfaces of constant magnetic flux in the outflow and can be
specified at the surface of the disk 
\citep[e.g.,][PP92]{Pelletier92}.   

{\bf Conservation of mass and magnetic flux} along a field line
can be combined into a single function
$k$ that is called the ``mass load'' of the wind
which is a constant along a magnetic field line; 
\begin{equation}
\rho {\bf v_p} = k {\bf B_p}.
\end{equation}
\noindent
This function represents the mass load per unit time,
per unit magnetic flux
of the wind. For axisymmetric flows, 
its value is preserved on each ring of  
field lines emanating from the accretion disk. 
Its value on each field line is determined
by physical conditions - including dissipative
processes - near the disk surface.
It may be more revealingly recast as
\begin{equation}
k(a) = { \rho v_p \over B_p} = {d \dot M_{\rm w} \over d \Psi},
\end{equation}
where $d\dot M_{\rm w}$ is the mass flow rate through an annulus
of cross-sectional area $dA$ through the wind and $d\Psi$
is the amount of poloidal magnetic flux threading through this same annulus.
The mass load profile is a function
of the footpoint radius $r_0$ of the wind on the disk. 

The toroidal field in rotating flows derives from the induction
equation;
\begin{equation}
B_{\phi} = {\rho  \over k} ( v_{\phi} - \Omega_0 r),
\end{equation}
where
$ \Omega_0$ is the angular
velocity of the disk at the mid-plane.
This result
shows that toroidal fields in the jet are formed by winding up the field from 
the source.  Their strength also 
depends on the mass loading as well as the jet density.
Denser winds should have stronger toroidal fields.
We note however, that the density does itself depend on the value 
of $k$.  Equation (9) also suggests that at 
higher mass loads, one has lower toroidal field strengths. 
This can be reconciled however, since it can be shown from the conservation laws
(see below) that the value of $k$ is related to the density of the outflow
at the Alfv\'en point on a field line; $k = (\rho_A/4\pi)^{1/2}$
(e.g., PP92).
Thus, higher mass loads correspond to denser winds and when
this is substituted into equation (9), we see that this also implies 
stronger toroidal fields.

{\bf Conservation of angular momentum} along each
field line leads to the conserved angular momentum per unit mass;
\begin{equation}
l(a) = r v_{\phi} - {r B_{\phi} \over 4 \pi k} = \mbox{const}. 
\end{equation}
The form for $l$ reveals that the total angular momentum
is carried by both the rotating gas (first term) as well
by the twisted field (second term), the relative proportion
being determined by the mass load.

The value of $l(a)$ that is  
transported along each field line is fixed by the
position of the Alfv\'en point in the flow, where the
poloidal flow speed reaches the Alfv\'en speed for the
first time
($m_{\rm A}=1$). It is easy to show that
the value of the specific angular momentum is;
\begin{equation}
l(a) = \Omega_0 r_{\rm A}^2 = (r_{\rm A}/r_0)^2 l_0.
\end{equation}
where 
$l_0 = v_{K,0}r_0 = \Omega_0 r_0^2$ is the 
specific angular momentum of a Keplerian disk.
For a field line starting at a point $r_0$ on the
rotor (disk in our case), the Alfv\'en radius is
$r_{\rm A}(r_0)$ and constitutes a lever arm for the flow.
The result shows that the angular momentum per unit
mass that is being extracted from the
disk by the outflow is a factor of $(r_{\rm A}/r_0)^2$ greater than
it is for gas in the disk.  For typical 
lever arms, one particle in the outflow can
carry the angular momentum of ten of its fellows left behind
in the disk.

{\bf Conservation of energy} along a field line
is expressed as a generalised version of Bernoulli's theorem
(this may be derived by taking the dot product of the
equation of motion with ${\bf B_p} $).  
Thus, there is a specific
energy $e(a)$ that is a constant along field lines, which
may be found in many papers (e.g., \ BP82 and PP92). 
Since the terminal speed $v_p = v_{\infty}$ of the disk
wind is much greater than its rotational speed, and
for cold flows, the pressure may also
be ignored, one finds the result: 
\begin{equation}
v_{\infty} \simeq 2^{1/2} \Omega_0 r_{\rm A} = (r_{\rm A}/r_0) v_{\rm esc,0}. 
\end{equation}

There are three important consequences for jet kinematics here;  
(i) that the terminal speed exceeds the {\it local} escape speed from 
its launch point on the disk by the lever arm ratio; 
(ii) the terminal speed scales with the Kepler speed as a function 
of radius, so that the flow will have an onion-like layering of
velocities, the largest inside, and the smallest on the larger scales,
as seen in the observations; and
(iii) that the terminal speed depends on the depth
of the local gravitational well at the footpoint of the
flow -- implying that it is essentially scalable to flows from disks around
YSOs of any mass and therefore universal.  

Another useful form of the conservation laws is the combination
of energy and angular momentum conservation to produce a new
constant along a field line (e.g., \ PP92);
$j(a) \equiv e(a) - \Omega_0  l(a) $.     
This expression has been used \citep{Anderson03} to 
deduce the rotation rate of the launch region on the Kepler
disk, where the observed jet rotation speed is $v_{\phi, \infty}$
at a radius $r_{\infty}$ and which is moving in the poloidal direction with 
a jet speed of $v_{p, \infty}$. 
Evaluating $j$ for a cold jet at infinity
and noting that its value (calculated at the foot point)
is $j(a_0)=-(3/2) v_{\rm K,0}^2$, one solves for the Kepler rotation at the 
point on the disk where this flow was launched: 
\begin{equation}
\Omega_0 \simeq v_{p,\infty}^2/ \left(2 v_{\phi,\infty} r_{\infty}\right). 
\end{equation}
When applied to the observed rotation of the 
Large Velocity Component (LVC) of the jet   
DG Tau \citep{Bacciotti02}, this yields a range of disk radii 
for the observed rotating material 
in the range of disk radii, 0.3--4 AU, and the magnetic  
lever arm is $r_{\rm A}/r_0 \simeq 1.8$--$2.6$.

\subsubsection{Angular Momentum Extraction}

How much angular momentum can such a wind extract from the disk? 
The angular momentum equation for the accretion disk undergoing
an external magnetic torque may be written:

\begin{equation}
\dot M_{\rm a} { d (r_0 v_0) \over dr_0} = - r_0^2 B_{\phi} B_z \vert_{r_0, H},
\end{equation}
\noindent
where we have ignored transport by MRI turbulence or spiral waves.
By using the relation between poloidal field and outflow on the 
one hand, as well as the link between the toroidal field and 
rotation of the disk on the other, the angular momentum equation 
for the disk yields one of the most profound scaling relations in disk wind
theory -- namely -- the link between disk accretion and mass outflow
rate (see KP00, PP92 for details):
\begin{equation}
\dot M_{\rm a} \simeq (r_{\rm A}/ r_0)^2 \dot M_{\rm w}.
\end{equation}
The observationally well known result that in many systems,
$\dot M_{\rm w} / \dot M_{\rm a} \simeq 0.1$ is a consequence of the 
fact that lever arms are often found in numerical and
theoretical work to be $r_{\rm A}/r_0 \simeq 3$ -- the observations
of DG Tau being a perfect example.

\subsubsection{Jet Power and Universality}

These results can be directly connected to 
the observations of momentum and energy transport in the molecular
outflows.
Consider the total mechanical power that is carried
by the jet, which may be written as \citep[e.g.,][]{Pudritz03};
\begin{equation}
L_{\rm jet}= {\textstyle{1 \over 2}}
\int_{r_{\rm i}}^{r_{\rm j}} d\dot M_{\rm w} v_{\infty}^2
\simeq {G M_* \dot M_{\rm a} \over 2r_{\rm i}}
\left(1 - {r_{\rm i}^2 \over r_{\rm j}^2}\right)
\simeq {\textstyle{1 \over 2}} L_{\rm acc} .
\end{equation}
This explains the observations of Class 0 outflows
wherein $L_{\rm w}/L_{\rm bol} \simeq 1/2$, 
since the main luminosity of the central source at this time 
is due to accretion and not nuclear reactions.
(The factor of $1/2$ arises from the dissipation
of some accretion energy as heat at the inner boundary).
The ratio of wind to stellar luminosity decreases at later 
stages because the accretion luminosity
becomes 
relatively small compared to the bolometric luminosity of the 
star as it nears the ZAMS. 

This result states that the wind luminosity taps the gravitational
energy release through accretion in the gravitational potential of
the central object -- and is a direct consequence of Bernoulli's
theorem.  This, and the previous results, imply 
that jets may be produced in any 
accreting system.  The lowest mass outflow that has yet
been observed corresponds to a proto-brown dwarf of
luminosity $\simeq 0.09 L_{\odot}$, a stellar 
mass of only $ 20 - 45 M_{Jup}$, 
and a very low mass disk $ < 10^{-4} M_{\odot}$ \citep{Bourke05}.

On very general grounds, disk winds are also likely to be active
during massive star formation \citep[e.g.,][]{Konigl99}.  Such
outflows may already start during the early collapse phase when the
central YSO still has only a fraction of a solar mass
\citep[e.g.,][]{Banerjee06, Banerjee07a}.  Such early
outflows may actually enhance the formation of massive stars via disk
accretion by punching a hole in the infalling envelop and releasing
the building radiation pressure \citep[e.g.,][]{Krumholz05}.

\subsubsection{Jet Collimation}

In the standard picture of hydromagnetic winds, collimation of an
outflow occurs because of the increasing toroidal magnetic field in
the flow resulting from the inertia of the gas.  Beyond the Alfv\'en
surface, equation~(8) shows that the ratio of the toroidal field to
the poloidal field in the jet is of the order $B_{\phi} / B_p \simeq
r/ r_{\rm A} \gg 1$, so that the field becomes highly toroidal.  In
this situation, collimation is achieved by the tension force
associated with the toroidal field which leads to a radially inwards
directed component of the Lorentz force (or ``$z$-pinch"); $ F_{\rm
Lorentz, r} \simeq j_z B_{\phi}$.  The stability of such systems is
examined in the next section.

In \citet{Heyvaerts89} it was shown
that two types of solution are possible depending upon
the asymptotic behaviour of the total current intensity in the jet;
\begin{equation}
 I = 
 2 \pi \int_0^r j_z(r',z')dr' = (c/2)
 r B_{\phi}. 
\end{equation}
In the limit that $I \rightarrow 0$ as
$r \rightarrow \infty $, the field lines are paraboloids
which fill space.  On the other hand, if the current
is finite in this limit, then the flow is collimated to cylinders.
The collimation of a jet therefore depends upon
its current distribution -- and hence on the radial distribution
of its toroidal field -- which, as we saw earlier,
depends on the mass load.  Mass loading therefore must play
a very important role in controlling jet collimation. 

It can be shown \citep[][PRO]{Pudritz06} from this that jets should
show different degrees of collimation depending on how they are mass
loaded.  As an example, neither the highly centrally concentrated,
magnetic field lines associated with the initial split-monopole
magnetic configuration used in simulations by \citet{Romanova97}, nor
the similar field structure invoked in the X-wind \citep[see review
by][]{Shu00} should become collimated in this picture.  On the other
hand, less centrally (radially) concentrated magnetic configurations
such as the potential configuration of \citet{Ouyed97} and BP82 should
collimate to cylinders.

This result also explains the range of collimation that is observed
for molecular outflows.  Models for observed outflows fall into two
general categories: the jet-driven bow shock picture, and a
wind-driven shell picture in which the molecular gas is driven by an
underlying wide-angle wind component such as given by the X-wind (see
review by \citet{Cabrit97}.  A survey of molecular outflows by
\citet{Lee00} found that both mechanisms are needed in order to
explain the full set of systems observed.

\section{Gravitational collapse, disks, and the origin of outflows}

\bef
\showtwo{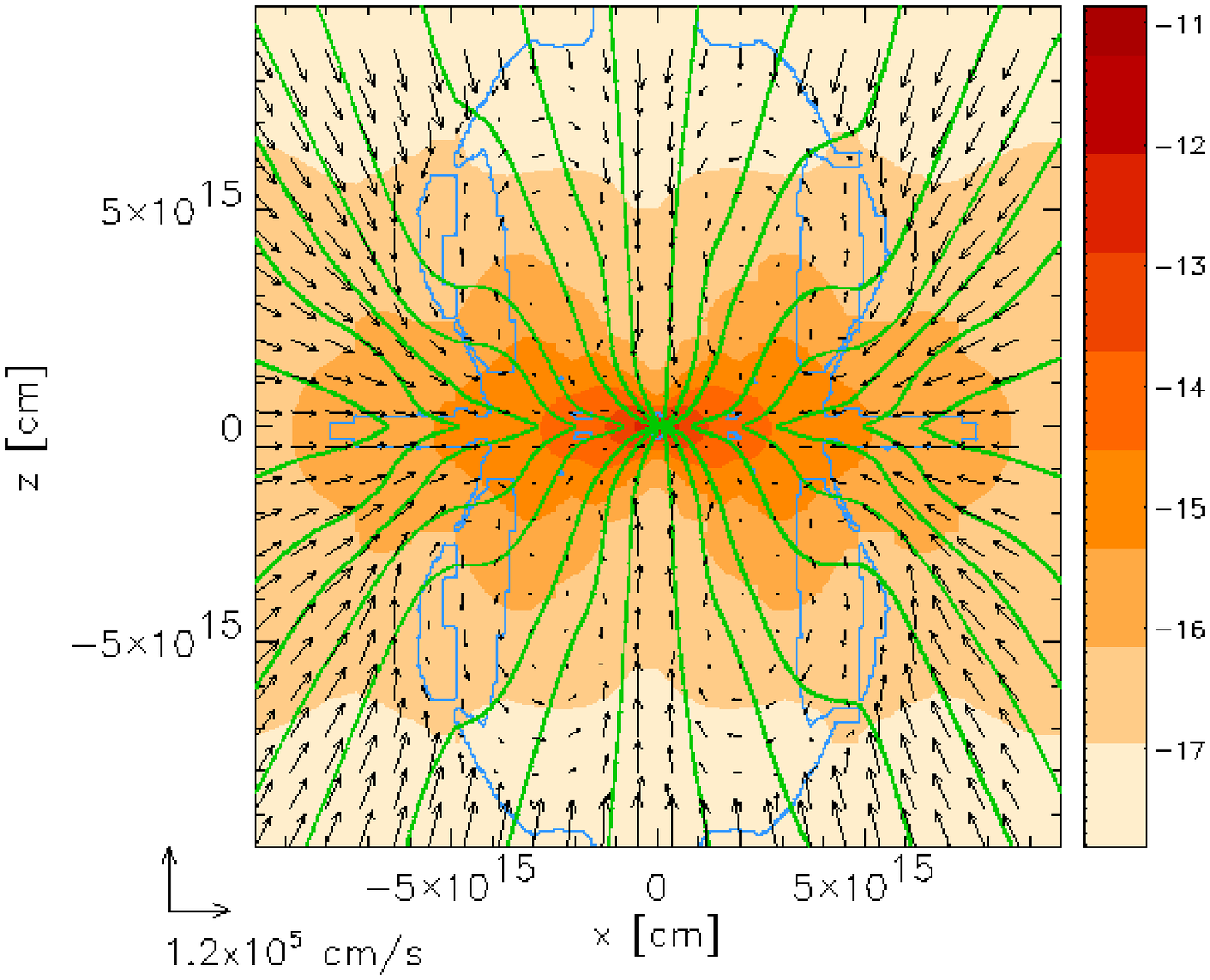}
	{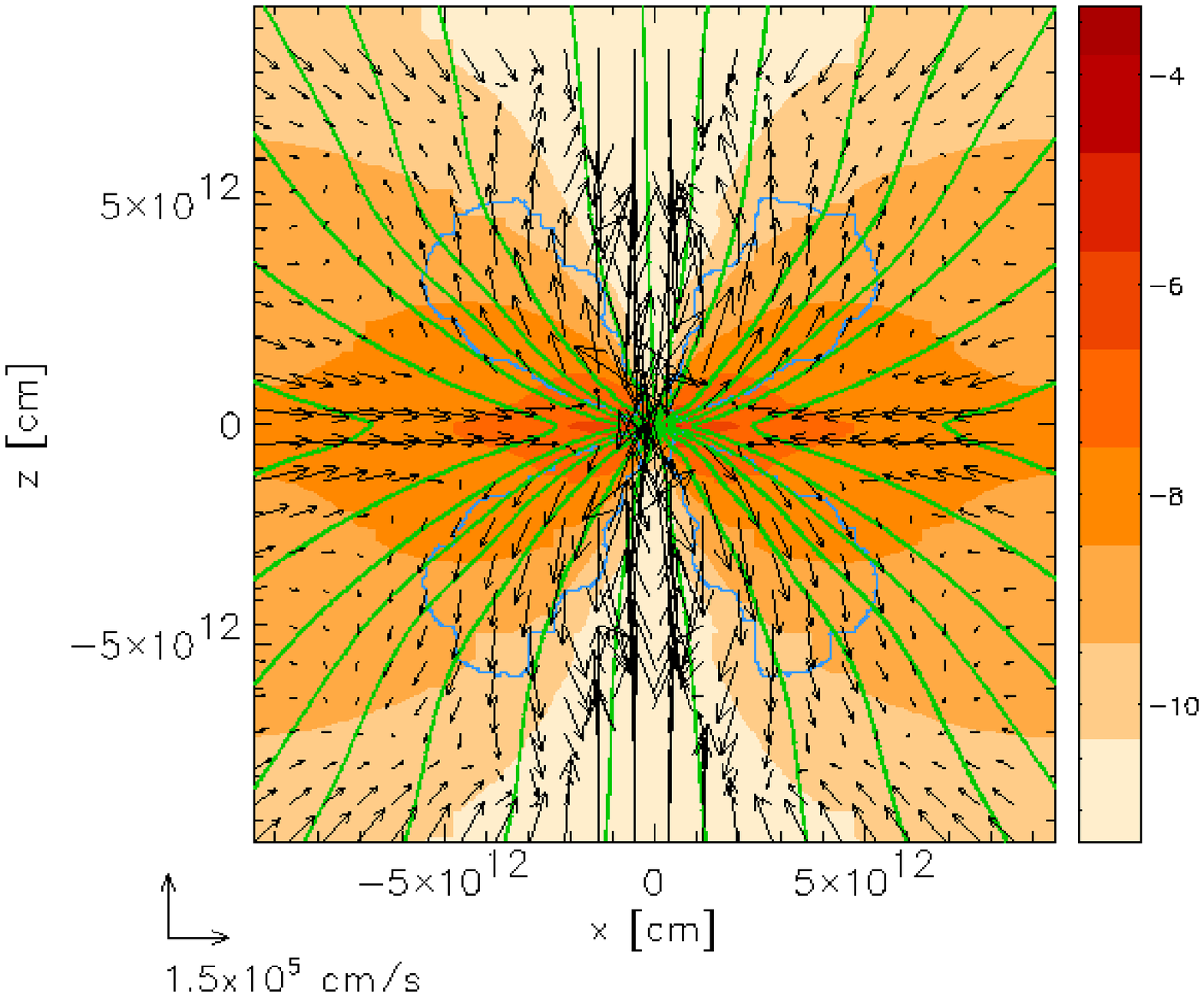}
\caption{Large scale outflow (left panel, scale of hundreds of AU), and
and small scale disk wind and jet formed (right panel, scale of a
fraction of an AU) during the gravitational collapse of a magnetised
B-E, rotating cloud core. Cross-sections through the disk and outflows
are shown -- the blue contour marks the Alf\'ven surface. Snapshots
taken of an Adaptive Mesh calculation at about 70,000 years into the
collapse.  [Adapted from \citet{Banerjee06}].}
\label{fig:outflow}
\eef

\bef
\showtwo{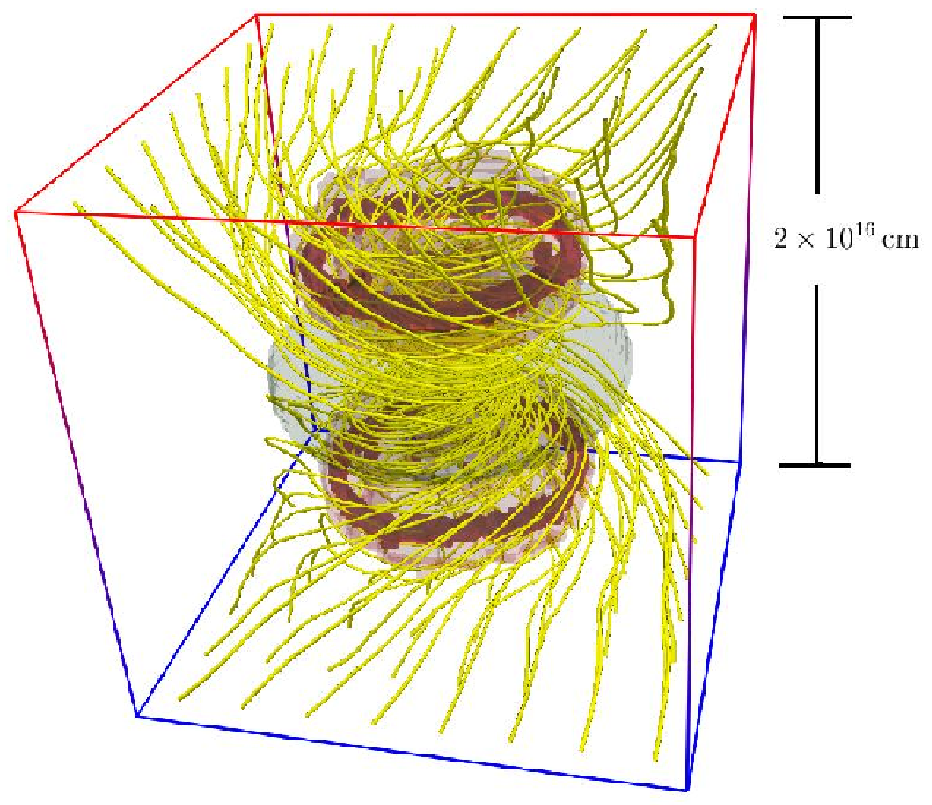}
	{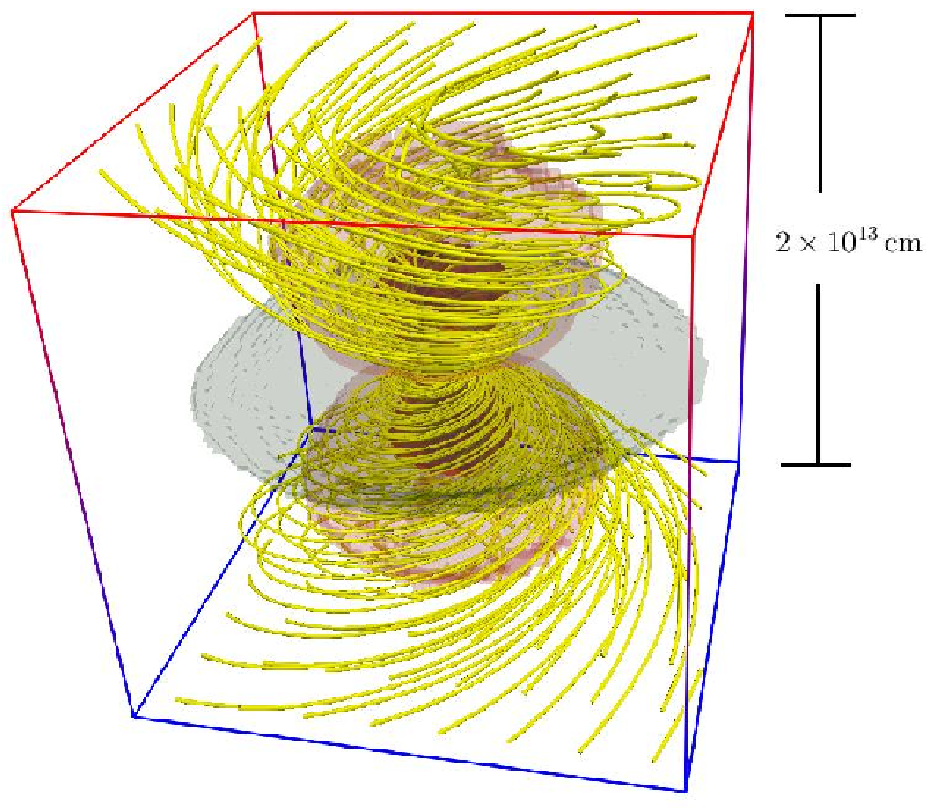}
\caption{Magnetic field line structure, outflow and disk. The two 3D
  images show the magnetic field lines, isosurfaces of the outflow
  velocities and isosurfaces of the disk structure at the end of our
  simulation ($t \simeq 7\times 10^{4} \, \ys$) at the two different
  scales as shown in Fig.~\ref{fig:outflow}. The isosurfaces of the
  upper panel refer to velocities $0.18\,\km\,\sec^{-1}$ (light red)
  and $0.34\,\km\,\sec^{-1}$ (red) and a density of $2\times 10^{-16}
  \, \g \, \cm^{-3}$ (gray) whereas the lower panel shows the
  isosurfaces with velocities $0.6\,\km\,\sec^{-1}$ (light red) and
  $2\,\km\,\sec^{-1}$ (red) and the density at $5.4\times 10^{-9} \,
  \g \, \cm^{-3}$ (gray).  [Adapted from \citet{Banerjee06}].}
\label{fig:fieldlines}
\eef

Jets are expected to be associated with gravitational collapse because
disks are the result of the collapse of rotating molecular cloud
cores.  One of the first simulations to show how jets arise during
gravitational collapse is the work of \citet{Tomisaka98, Tomisaka02}.
These authors used as initial conditions the collapse of a magnetised
rotating filament (cylinder) of gas and showed that this gave rise to
the formation of a disk from which a centrifugally driven disk wind
was produced.

Banerjee \& Pudritz revisited the problem of collapsing magnetised
molecular cloud cores in ~\citet{Banerjee06} (low mass cores, BP06)
and \citet{Banerjee07a} (high mass cores, BP07). Here the initial
cores are modelled as supercritical Bonnor-Ebert (B-E) spheres with a
slight spin. B-E spheres are well controlled initial setups for
numerical experiments and there are plenty of observed cores which can
fit by a B-E profile~\citep[see][on a compilation of hydrostatic
cores]{Lada07}. Recent efforts have focused on understanding the
evolution of magnetised B-E spheres~\citep[e.g.][]{Matsumoto03,
Machida05a, Machida05b, Hennebelle07b, Hennebelle08}.  Whereas purely
hydrodynamic collapses of such objects never show outflows
\citep[e.g.,][]{Foster93, Banerjee04}, the addition of a magnetic
field produces them.

Outflows are also implicated in playing a fundamental role in
controlling the overall amount of mass that is incorporated into a new
star from the original core. Observations of the so-called core mass
function (CMF) that describes the spectrum of core masses, show that
it has nearly the identical shape as the initial mass function (IMF)
that describes the spectrum of initial stellar masses - with one
caveat.  The CMF needs to be muliplied by a factor of 1/3 \citep[in one well
studied case - see][]{Alves07} in order to align them.  The
suggestion is that protostellar outflows are responsible for removing
as much as 2/3 of a core's mass in the process of collapse and star
formation.

In BP06, the FLASH AMR code~\citep{FLASH00} was used to follow the
gravitational collapse of magnetised molecular cloud cores.  This code
allowed the Jeans length to be resolved with at least 8 grid points
throughout the collapse calculations. This ensures compliance with the
Truelove criterion~\citep{Truelove97} to prevent artificial
fragmentation.

The results of the low mass core simulation are shown in
fig.~\ref{fig:outflow} and fig.~\ref{fig:fieldlines}.  they show the
end state (at about 70,000 yrs) of the collapse of a B-E sphere that
is chosen to be precisely the Bok globule observed by \citet{Alves01}
-- whose mass is $2.1 M_{\odot}$ and radius $R=1.25 \times 10^4$ AU at
an initial temperature of 16 K.  Two types of outflow can be seen: (i)
an outflow that originates at scale of $ \simeq 130$ AU on the forming
disk that consists of a wound up column of toroidal magnetic field
whose pressure gradient pushes out a slow outflow; and (ii) a
disk-wind that collimates into a jet on scale of 0.07 AU.  A tight
proto-binary system has formed in this simulation, whose masses are
still very small $\le 10^{-2} M_{\odot}$, which is much less than the
mass of the disk at this time $\simeq 10^{-1} M_{\odot}$.  The outer
flow bears the hallmark of a magnetic tower, first observed by
\citet{Uchida85b}, and studied analytically by \citet{LyndenBell03}.
Both flow components are unbound, with the disk wind reaching $3$ km
s$^{-1}$ at 0.4 AU which is quite super-Alfv\'enic and above the local
escape speed.  The outflow and jet speeds will increase as the central
mass grows.

\bef
\showtwo{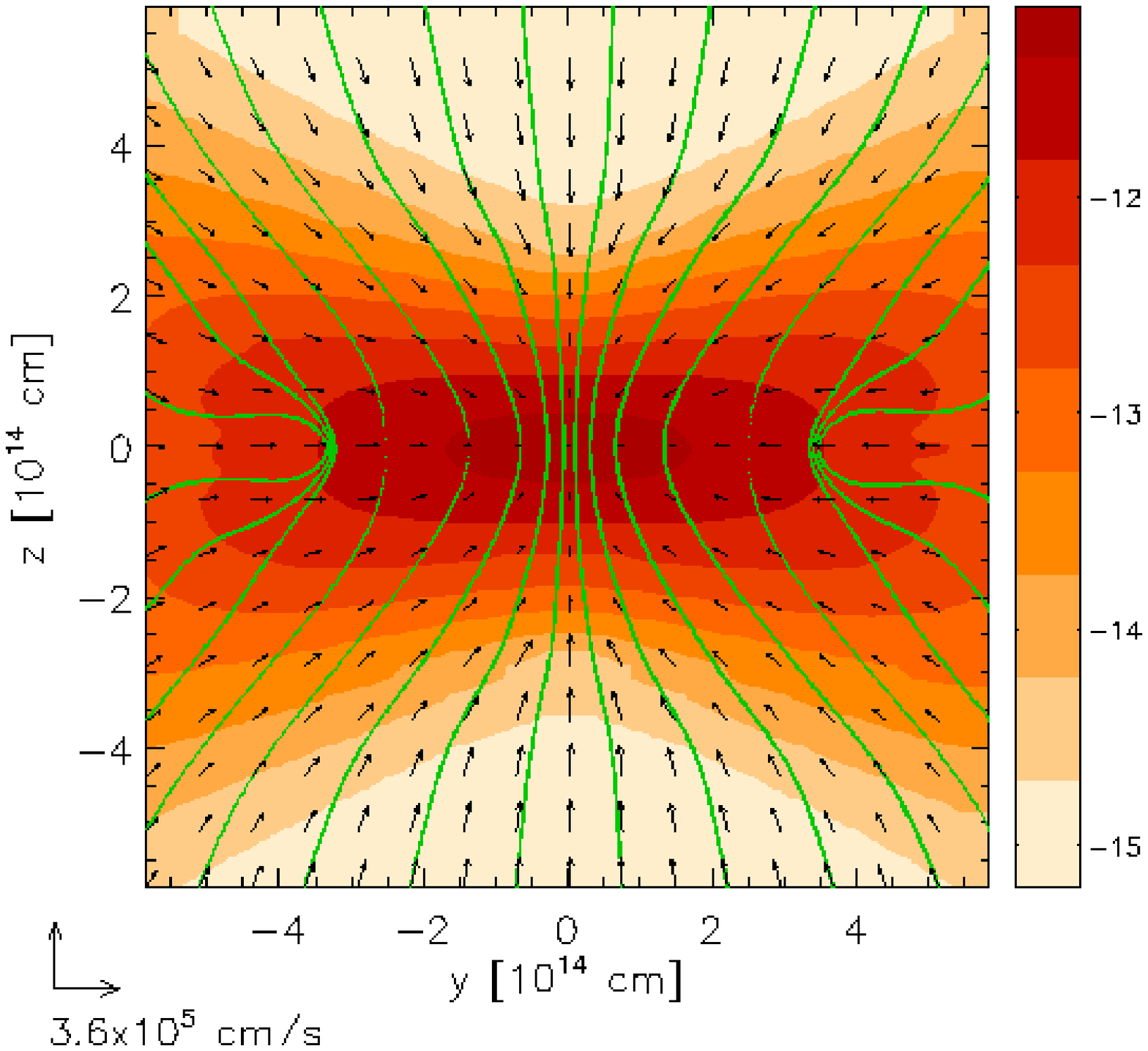}
	{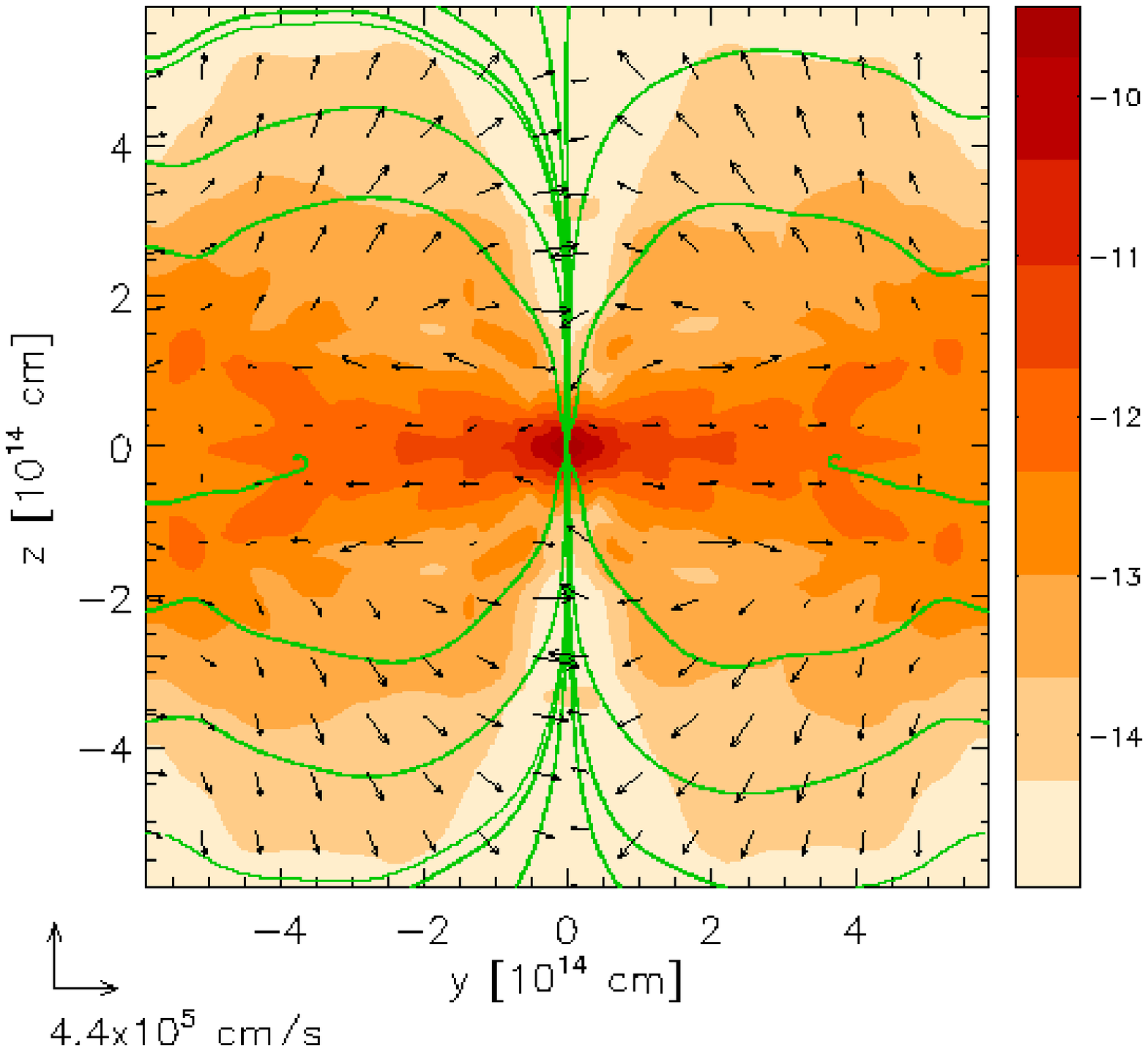}
\caption{Snapshots of the central region of a collapsing magnetised
  massive cloud core ($M_{\sm{core}} \sim 170 \, \Msol$). The left
  panel shows the situation at $t = 1.45\times 10^4 \, \ys$ ($1.08 \,
  t_{\sm{ff}}$) into the collapse and before the flow reversal and
  right panel shows the configuration $188 \, \ys$ later when the
  outflow is clearly visible. [Adapted from \citet{Banerjee07a}].}
\label{fig:massive_outflow}
\eef

In \citet{Banerjee07a} a similar setup to the one described above is
used to study the formation of massive stars.  In an unified picture
of star formation one would also assume that outflows and jets will
also be launched around young massive stars. Due to the deeply
embedded nature of massive star formation observations of jets and
outflows are much more difficult than for the more isolated low mass
stars. Nevertheless, there is increasing observational evidence for
outflows around young massive stars \citep[e.g.,][]{Arce07, Zhang07}.

The simulations of collapsing magnetised massive cores ($\sim 170 \,
\Msol$) show clear signs of outflows during the early stages of
massive star formation. Fig.~\ref{fig:massive_outflow} shows two 2D
snap-shots from this simulations where the onset of an outflow is
visible. Early outflows of this kind might have a large impact on the
accretion history of the young massive star. \citet{Krumholz05} showed
that cavities blown by outflows help to release the radiation pressure
from the newly born massive star which in turn relaxes radiation
pressure limited accretion onto the central object.

From these simulations one can conclude that the theory and computation
of jets and outflows is in excellent agreement with new observations
of many kinds.  Disk winds are triggered during magnetised collapse
and persist throughout the evolution of the disk.  They efficiently
tap accretion power, transport a significant part portion of the
disk's angular momentum, and can achieve different degrees of
collimation depending on their mass loading.

\section{Feedback from collimated protostellar jets?}

The interstellar medium (ISM) and star forming molecular clouds are
permeated by turbulent, supersonic gas motions~\citep[e.g.~see recent
reviews][and references therein]{Elmegreen04, MacLow04,
Ballesteros07}. Supersonic turbulence in molecular clouds is known to
play at least two important roles: it can provide pressure support to
help support molecular clouds against rapid collapse, and it can also
produce the system of shocks and compressions throughout such clouds
which fragment the cloud into the dense cores that are the actual
sites of gravitational collapse and star formation.  One of the major
debates in the literature is on the question of just how long this
turbulence can be sustained - and with it, star formation.  Despite
the importance of supersonic turbulence for the process of star
formation its origin is still unclear.  \citet{Norman80} proposed that
Herbig-Haro (HH) outflows could provide the energetics to drive the
turbulence in molecular clouds that could keep star formation active
for a number of cloud free-fall times (i.e., several million
years). This is an attractive idea as this process could lead to a
self-regulating star formation environment~\citep[see also][]{Li06,
Nakamura07}.

In a recent investigation by \citet{Banerjee07b} this idea was
addressed within a detailed study of feedback from collimated jets on
their supra-core environment. These studies show that supersonic
turbulence excited by collimated jets decays very quickly and does not
spread far from the driving source, i.e. the jet. Supersonic
fluctuations, unlike subsonic ones, are highly compressive. Therefore,
the energy deposited by a local source, like a collimated jet, stays
localised as the compressed gas either heats up (in the non-radiative
case) or is radiated by cooling processes. The re-expansion of the
compressed regions excites only subsonic or marginally supersonic
fluctuations. This is contrary to almost incompressible subsonic
fluctuations. These fluctuations travel like linear waves with little
damping into the ambient medium and make up most of the overall
velocity excitations.

In the study of \citet{Banerjee07b} a series of numerical experiments
were performed with the adaptive mesh refinement (AMR) code
FLASH. Here the jet is modelled as a kinetic energy injection from the
box boundary. The energy injection could be switched on and off after
a certain amount of time, i.e. the jets are either continuously driven
or transient. These jets are either interacting with a homogeneous or
clumpy environment, where the latter is modelled as a spherical
overdensity. Additionally the influence of magnetic fields on
jet-excited fluctuations where considered in this study.

The authors quantify the jet-excited motions of the gas using velocity
probability distribution functions (PDF) which can be regarded as
volume weighted histograms. From these PDF the distinction between
subsonic and supersonic fluctuations comes about naturally: the
supersonic regime is strongly suppressed compared to the subsonic
regime. Additionally, separate decay laws of the kinetic energy for
the subsonic and supersonic regimes where derived to quantify their
different behaviours with time.

\bef
\showone{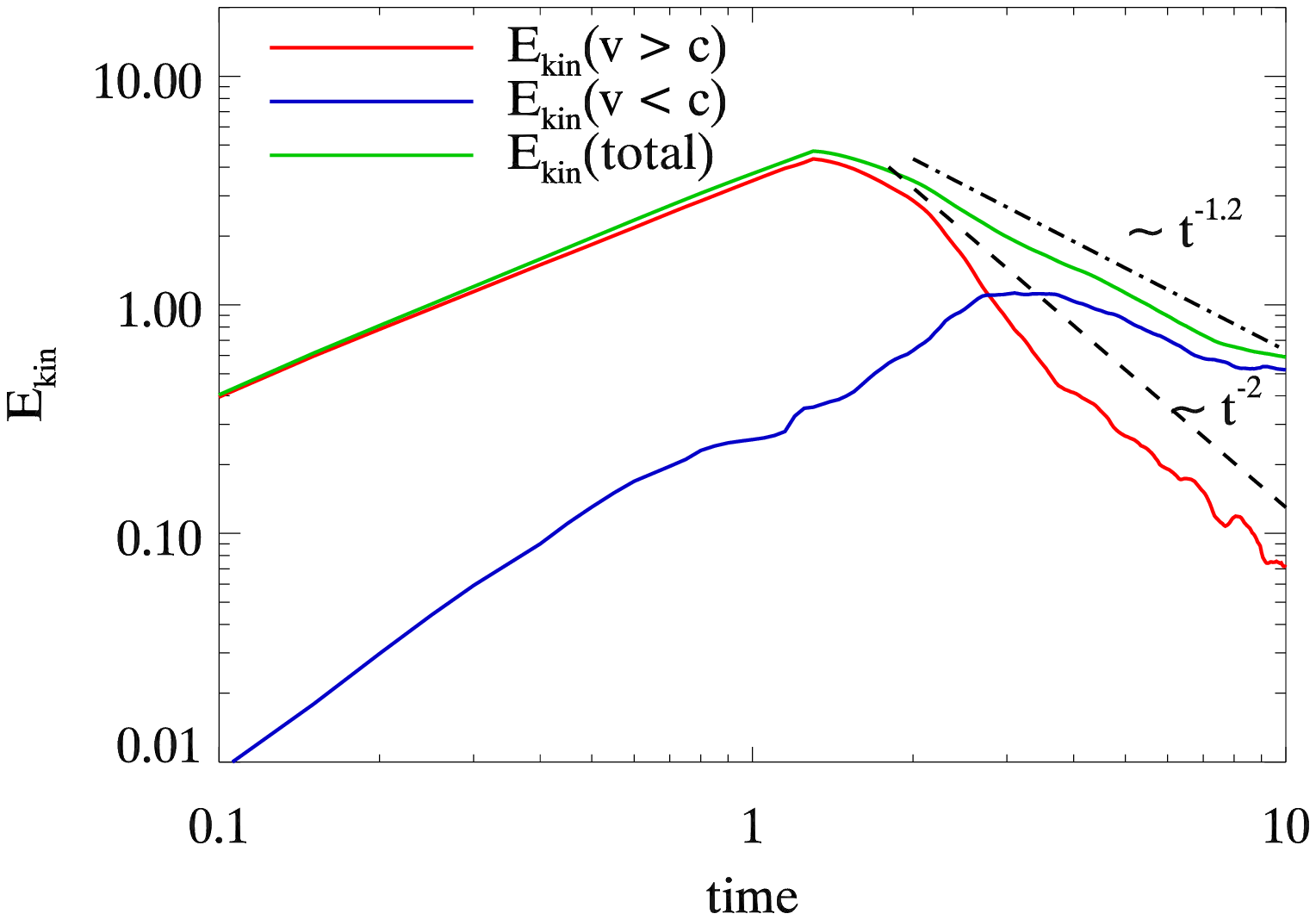}
\caption{Time evolution of the kinetic energies
  ,$E_{\sm{kin}}$, for a transient jet with a speed of Mach 5. The
  quantities are divided into a supersonic regime, $v > c$, and a
  subsonic regime, $v < c$. The decay of supersonic energy
  contributions is much faster (faster than $\propto t^{-2}$) than the
  subsonic one. The time shift between the peaks in the kinetic
  energies shows that the subsonic fluctuations are powered by the
  decay of the supersonic motions. [Adapted from \citet{Banerjee07b}]}
\label{fig:Ekin}
\eef

\bef
\showtwo{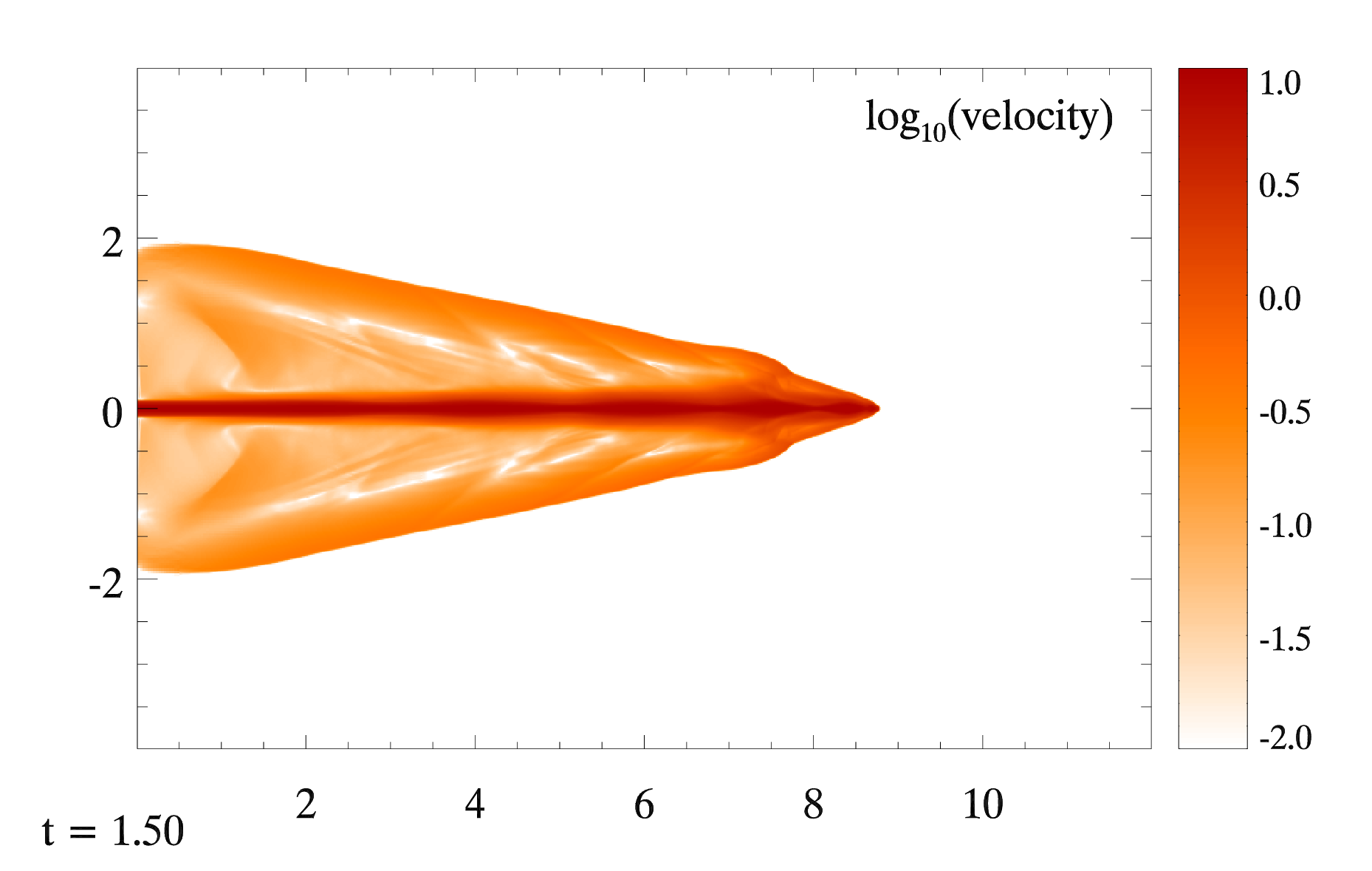}
	{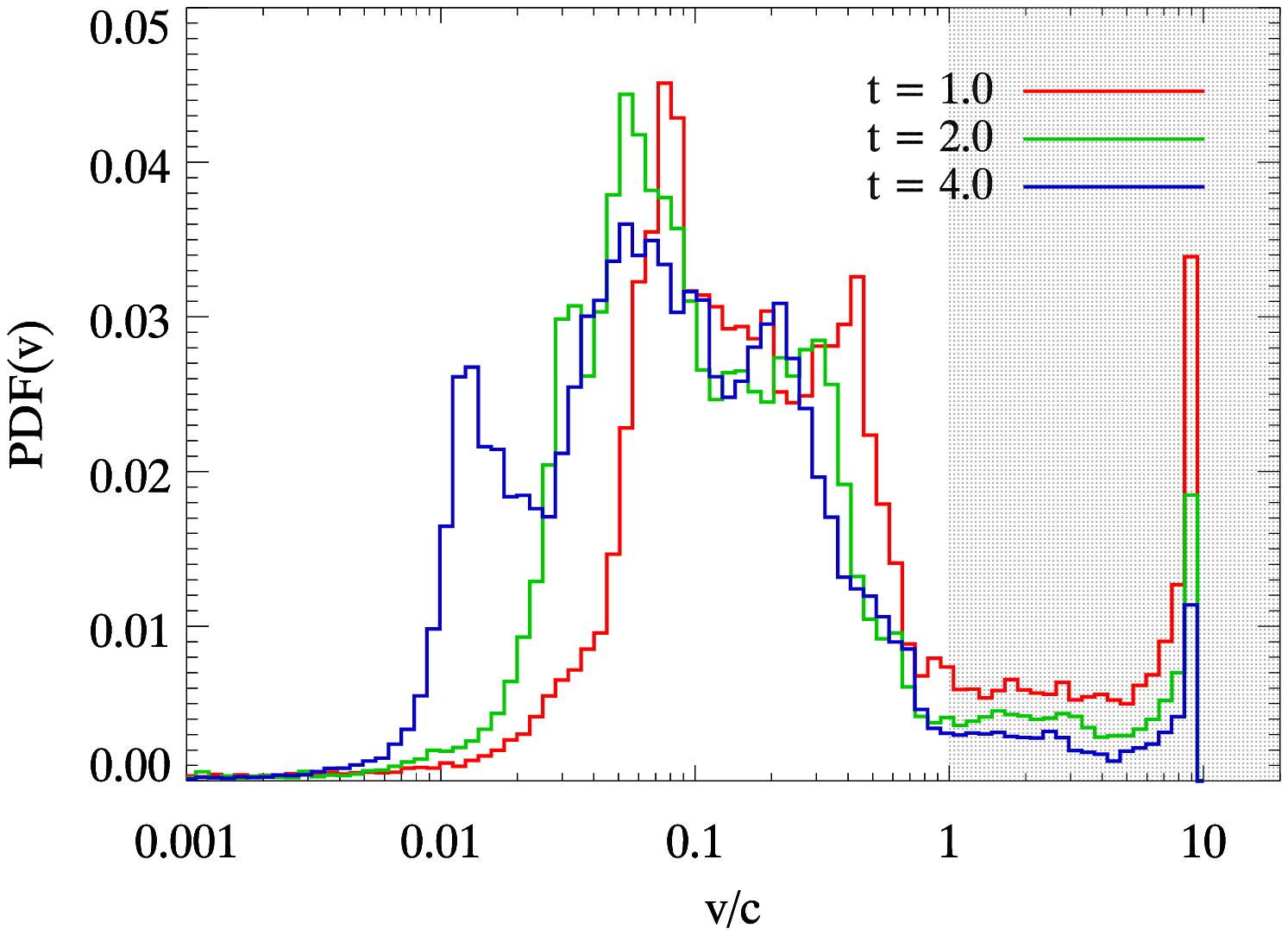}
\caption{Velocity structure and velocity PDFs at different times of a
  continuously driven Mach 10 jet. Essentially no supersonic
  fluctuations in the ambient gas get excited by the jet (the peak at
  $v/c = 10$ is the jet itself). [Adapted from \citet{Banerjee07b}]}
\label{fig:JetMach10}
\eef

Fig.~\ref{fig:Ekin} shows that the kinetic energy of the supersonic
fluctuations decays much faster than subsonic
contributions. Furthermore, one can see from fig.~\ref{fig:JetMach10}
that the jet-excited fluctuations are mainly subsonic which is due to
the fact that the supersonic excitations do not travel far from the
edge of the jet (see right panel of fig.~\ref{fig:JetMach10}).  

This study shows that supersonic fluctuations are damped quickly
because they excite mainly compressive modes. The re-expansion of the
compressed overdensities drives mainly subsonic velocity fluctuations
that then propagate further into the ambient medium. This is in spite
of the appearance of bow shocks and instabilities. In particular,
instabilities such as Kelvin-Helmholtz modes at the edge of the jet,
develop most efficiently for transonic or slower
velocities. High-velocity jets, on the contrary, are bullet-like and
stay very collimated, transiting the surrounding cloud without
entraining much of its gas.  From the point of view of jet-driven
supersonic turbulence in molecular clouds this is a dilemma which is
difficult to circumvent. Even in the case of overdense jets which
affect more gas of the surrounding media and have higher momenta, the
supersonic motions do not propagate far from their source.

Simulations of magnetised jets in this study show that jets stay
naturally more collimated if the magnetic field is aligned with the
jet axis and therefore entrain less gas. Furthermore, perpendicular
motions are damped by magnetic tension preventing a large spread of
high amplitude fluctuations. On the other hand, perpendicular field
configurations support the propagation of such modes which are able to
spread into a large volume. Nevertheless, the vast majority of these
motions are still subsonic.

This study shows that collimated jets from young stellar objects are
unlikely drivers of large-scale {\em supersonic} turbulence in
molecular clouds. Alternatively it could be powered by large scale
flows which might be responsible for the formation of the cloud
itself~\citep[e.g.,][]{Ballesteros99b}. Energy cascading down from the
driving scale to the dissipation scale will then produce turbulent
density and velocity structure in the inertial range in between
\citep[][]{Lesieur97}.  If the large-scale dynamics of the
interstellar medium is driven by gravity \citep[as suggested, e.g.,
by][]{LiMacLowKlessen05,LiMacLowKlessen06} gravitational contraction
would also determine to a large extent the internal velocity structure
of the cloud. Otherwise, blast waves and expanding shells from super
novae are also viable candidates to power supersonic turbulence in
molecular clouds~\cite[see e.g.,][]{MacLow04}.

\section{Relativistic jets: theory}

The theory of relativistic MHD jets is best described in the SRMHD
regime.  The essential parameter is the {\em magnetisation} parameter
\citep{Michel69},
\begin{equation}
\sigma = \frac{\Psi^2\Omega_F^2}{4 \dot{M} c^3}.
\label{eq_sigdef}
\end{equation}
The iso-rotation parameter $\Omega_F$ is frequently interpreted as the
angular velocity of the magnetic field lines.
The function $\Psi = B_p r^2$ is a measure of the magnetic field distribution
(see Li 1993), and $\dot{M}\equiv \pi\rho v_p R^2$
is the mass flow rate within the flux surface.
Equation ({\ref{eq_sigdef}) demonstrates that the launch of a highly
relativistic (i.e. highly magnetised) jet essentially requires at
least one of three conditions --  rapid rotation, strong magnetic
field and/or a comparatively low mass load.

In the case of a spherical outflow ($\Psi = const$) with negligible gas
pressure one may derive the Michel scaling  between the asymptotic Lorentz
factor and the flow magnetisation \citep{Michel69},
\begin{equation}
\Gamma_{\infty} =
\sigma^{1/3}
\label{eq_sigmich}
\end{equation}
Depending on the exact magnetic field distribution $\Psi(r,z)$ (which
describes the {\em opening} of the magnetic flux surfaces), in a {\em
collimating jet} the matter can be substantially accelerated beyond
the fast point magnetosonic point \citep{Begelman94, Fendt96}, as it
is moved from infinity to a finite radius of several Alfv\'en radii.
As a result, the power law index in Eq.\,({\ref{eq_sigmich}) can be
different from the Michel-scaling \citep[see][]{Fendt96, Vlahakis01,
Vlahakis03b}.

The {\em light cylinder} (hereafter l.c.)  is located at the
cylindrical radius $r_l = c/\Omega_F$.  At the l.c. the velocity of
the magnetic field lines ``rotating'' with angular velocity $\Omega_F$
coincides with the speed of light\footnote{Outside the l.c. the
magnetic field lines ``rotate'' faster than the speed of light. As the
field line is not a physical object, the laws of physics are not
violated.}.  The l.c. has to be interpreted as the Alfv\'en surface in
the limit of vanishing matter density (force-free limit).  {\it The
location of the l.c. determines the relativistic character of the
magnetosphere. If the light cylinder is comparable to the dimensions
of the object investigated, a relativistic treatment of MHD is
required}.

While SRMHD can be invoked to study jets at scales larger than the
gravitational radius of the black hole, close to the horizon {\em
general relativity} becomes relevant and the MHD equations need to be
coupled to General Relativity. This often requires the use of
sophisticated numerical codes in order to capture the complexity of
jet physics in GRMHD.  In the (non-rotating) black hole's Schwarzschild
spacetime the GRMHD equations are identical to the SRMHD equations in
general coordinates, except for the gravitational force terms and the
geometric factors of the lapse function.

\bef
\centering
\showonevert{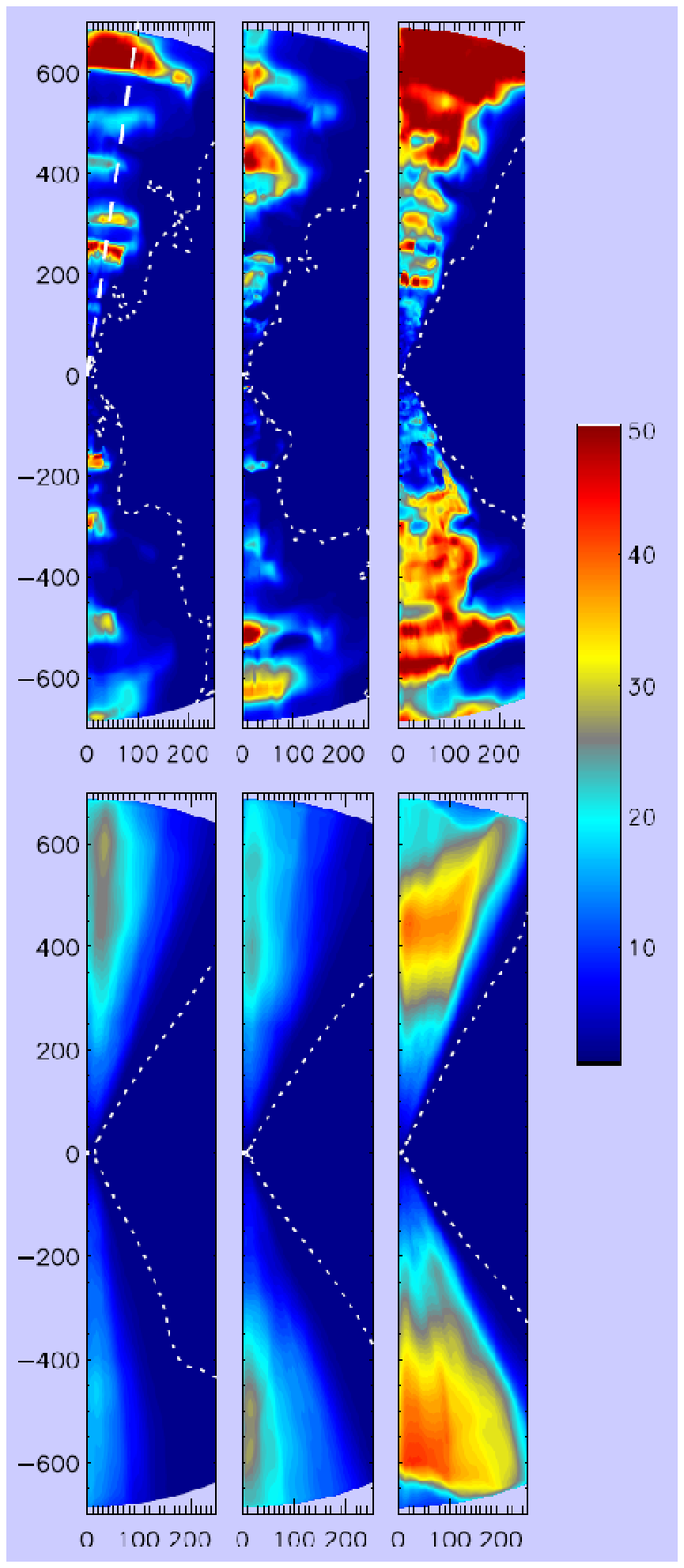}
\caption{ Lorentz factor from GRMHD simulations; adopted from
\citet{DeVilliers05}. In all panels, the axes are in units of $M$
(here $1\,M \approx 4$ km). The black hole is located at the origin.
The top row, left to right, shows plots of Lorentz factor for
decreasing black hole rotation $a/M=0.995, 0.9, 0.0$,
respectively. The dotted contour marks the boundary of the jets. The
only region where elevated values of $\Gamma$ are found is in the jets
(and also in the bound plunging inflow near the black hole, which is
not resolved at the scale of this figure).  Maximum values of $\Gamma$
are found in knots that appear episodically in the upper and lower
parts of the funnel. The bottom three panels show, from left to right,
the corresponding time-averaged value of the Lorentz factor.  The
plots show evidence of spin-dependent collimation: cylindrical
collimation is seen in the high-spin models, while the zero-spin model
shows no such collimation. These plots also show evidence of an
extended acceleration zone: large Lorentz factors are built up over
the full radial range.  }
\label{fig:ultra-jets1}
\eef

\bef
\showone{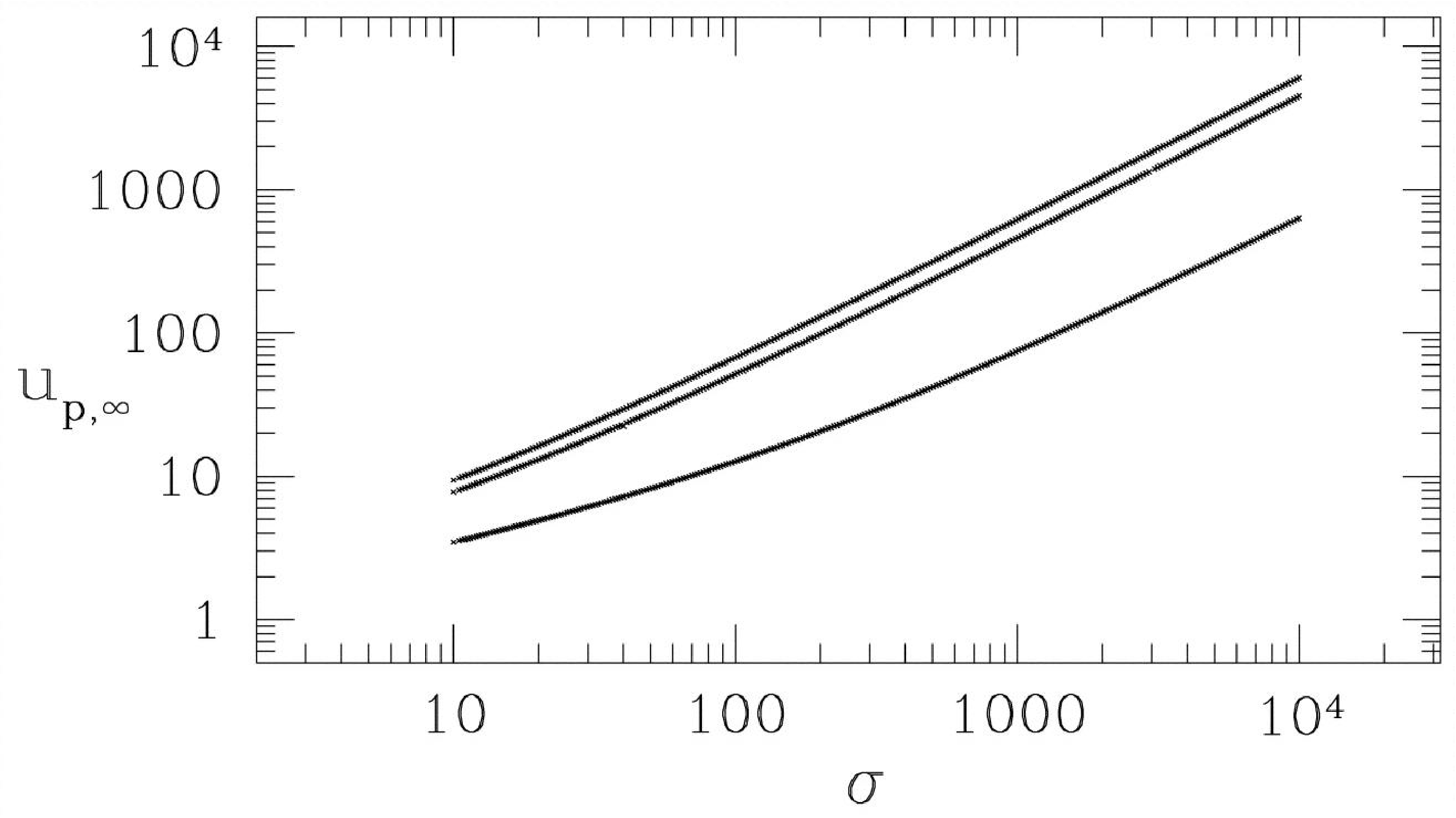}
\caption{ Asymptotic Lorentz factor ($u_{p,\infty}$) from SRMHD
 simulations, for different magnetisation parameter $\sigma$
 \citep[from][]{Fendt04}. Lorentz values as high as 5000 can be
 reached for strongly magnetised sources.  }
\label{fig:ultra-jets2}
\eef

\section{SRMHD and GRMHD Simulations}

Full GRMHD numerical simulations on the formation of jets near a black
hole were first been performed by \citet{Kudoh98}.  This was followed
by a plethora of GRMHD codes with fixed spacetimes used to investigate
jets from accreting black holes \citep[e.g.,][]{DeVilliers03,
Gammie03, Komissarov04, Anton06, Anninos05}.  Here we focus on the
most recent simulations that have managed to reach the highest Lorentz
factors for the study of GRMHD jets.

Recent state-of-the-art GRMHD simulations have shown that the
accretion flow launches energetic jets in the axial funnel region
(i.e.  low density region) of the disk/jet system, as well as a
substantial coronal wind \citep{DeVilliers05}. The jets feature
knot-like structures of extremely hot, ultra-relativistic gas; the gas
in these knots begins at moderate velocities near the inner engine,
and is accelerated to ultra-relativistic velocities (Lorentz factors
of 50, and higher) by the Lorentz force in the axial funnel. The
increase in jet velocity takes place in an acceleration zone extending
to at least a few hundred gravitational radii from the inner
engine. The overall energetics of the jets are strongly
spin-dependent, with high-spin black holes producing the highest
energy and mass fluxes. In addition, with high-spin black holes, the
ultra-relativistic outflow is cylindrically collimated within a few
hundred gravitational radii of the black hole, whereas in the
zero-spin case the jet retains a constant opening angle of
approximately 20 degrees.

Figure \ref{fig:ultra-jets2} shows the Lorentz factor, $\Gamma$, in
 the funnel outflow from the GRMHD simulations discussed in
 \citet{DeVilliers05}. Elevated values of the Lorentz factor are found
 in compact, hot, evacuated knots that ascend the funnel radially; the
 combination of low density and high temperature in the
 knots\footnote{\citet{Ouyed97} discuss the appearance of knots in
 MHD jet simulations; the knots seen in those simulations consisted of
 high density material, in contrast to what is observed in the present
 simulations.}. The highest values of Lorentz factor reach the maximum
 allowed by the code ($\sim 50$, and test runs show that much higher
 values could be reached with a higher ceiling), and these are only
 found at large radii, suggesting the presence of an extensive region
 where the knots are gradually accelerated to higher Lorentz factors.

While the details of the mechanism by which GRMHD jets are launched are
yet to be fully understood, the following summary addresses the
salient points of the complex dynamics by which unbound the plasma
from the vicinity of the compact star \citep[see discussion
in][]{DeVilliers05, McKinney07}: In the region of the accretion flow
near the marginally stable orbit, both pressure gradients and the
Lorentz force act to lift material away from the equatorial
plane. Some of this material is launched magneto-centrifugally in a
manner reminiscent of the scenario of \citet{Blandford82}, generating
the coronal wind; some of this material, which has too much angular
momentum to penetrate the centrifugal barrier, also becomes part of
the massive funnel-wall jet. There is also evidence that the
low-angular momentum funnel outflow originates deeper in the accretion
flow. Some of this material is produced in a gravitohydromagnetic
interaction in the ergosphere \citep{Punsly90}, and possibly in a
process similar to that proposed by \citet{Blandford77} where
conditions in the ergosphere approach the force-free limit. The
material in the funnel outflow is accelerated by a relatively strong,
predominantly radial Lorentz force; gas pressure gradients in the
funnel do not contribute significantly.

In the context of GRBs, the Lorentz factor $\Gamma$ of the
relativistic wind must reach high values ($\Gamma\sim 10^2-10^4$) both
to produce $\gamma$-rays and to avoid photon-photon annihilation along
the line of sight, whose signature is not observed in the spectra of
GRBs \citep{Goodman86}.  As we have seen, GRMHD simulations indicate
Lorentz factors of up to $\sim 50$ close to the black hole with
indication of higher $\Gamma$ far beyond the central object where
SRMHD treatment is sufficient.  The extreme Lorentz factors have been
confirmed by recent SRMHD simulations which have found solutions with
$\gamma$ as high as $\simeq 5000$ as shown in Figure
\ref{fig:ultra-jets2} \citep[see][for more discussion]{Fendt04}.

\section{Non-relativistic versus relativistic MHD jets}\label{sec:diffs}

While SRMHD and GRMHD jet simulations show similarities with
 non-relativistic regime, there are nevertheless some important
 differences in the underlying physics.

\begin{itemize}
\item  The existence of the l.c. as a natural length scale in relativistic MHD is
not consistent with the assumption of a {\em self-similar} jet structure (as is 
 often assumed in non-relativistic MHD jet models).  
The latter holds even more when general relativistic effects are considered.
\item Contrary to Newtonian MHD, in the relativistic case {\em electric fields}
cannot be neglected.
The poloidal electric field component is directed perpendicular to
the magnetic flux surface. Its strength scales with the l.c. radius,
$E_{\rm p} = E_{\perp} = (r/r_l) B_{\rm p}$.
As a consequence of $E_{\rm p} \simeq B_{\rm p}$, the effective magnetic
pressure can be lowered by a substantial amount \citep{Begelman94}.
\item In relativistic  MHD 
the poloidal Alfv\'en speed $u_{\rm A}$ becomes complex for $r > r_l$,
$u_{\rm A}^2 \sim B_p^2\,(1 - (r/r_l)^2)
= b_p^2-E_{\perp}^2$\,.
Therefore, Alfv\'en waves cannot propagate beyond the l.c. and only fast
magnetosonic waves are able to exchange information across the jet.

\item In the relativistic case, break-up of the MHD approximation is an
issue.  The problem is hidden in the fact that one may find
arbitrarily high velocities for an arbitrarily high flow
magnetisation.  However, an arbitrarily high magnetisation may be in
conflict with the intrinsic {\em MHD condition} which requires a
sufficient density of charged particles in order to be able to drive
the electric current system \citep{Michel69}.  Below the
Goldreich-Julian particle density $n_{\rm G}$ \citep{Goldreich69} the
concept of MHD as applied to relativistic jet breaks down.
\end{itemize}

The points above imply that with due regard to the break down of the MHD
approximation, it is feasible to scale the physics of jets from non-relativistic to
ultra-relativistic MHD regime. GRMHD simulations suggest  that a continuous scaling is
more likely for slowly rotating black holes  (where the radius of
marginal stability is rather large and in a comparatively low-gravity
region of spacetime)  but show that a full general relativistic
electrodynamic treatment is required to robustly treat the case of
rapidly rotating black holes.  GRMHD simulations also show that the
situation is even more complex in the  ``plunge region". This is the
region of the disk within the radius of marginal stability in which the
accretion flow is undergoing rapid inwards acceleration (ultimately
crossing the event horizon at the velocity of light as seen by a
locally non-rotating observer).   Unless the magnetic field is
extremely strong, this is a region where inertial forces will dominate
and the commonly employed force-free approximation breaks down.  As
a particle gets closer to the central object it crosses three regions
defining the MHD, force-free and inertial regimes, respectively.   
Another crucial aspect of the physics inherent to ultra-relativistic jets emanating
from the near vicinity of rapidly rotating central objects is the role
 played by the dynamically important electric field
 \citep[e.g.,][]{Lyutikov07, Ouyed07} in driving plasma instabilities
generally leading to pair creation. Pair-creation and subsequent
 annihilation into radiation has yet to be taken into account, at least
  consistently, in GRMHD or/and GREMD codes which are mostly based
 on mass conservation schemes. 

In summary, it appears that a universal aspect of all jets is their ability to tap
their energy from the accretion of gas into the gravitational potential of the underlying central
object.  Jets of
all stripes  may also share common morphological features far beyond the source.  However, complete unification of
non-relativistic with relativistic regimes for jets probably breaks down close to the central
black holes, where jet physics will also depend on the rotation of the hole and the production
of relativistic plasmas.

\noindent
\textbf{Acknowledgements }: We thank the organisers for the
opportunity to present this work in such a stimulating conference, in
such a spectacular setting. Some material in this chapter ( \S 1.3) has
appeared, in different form in another recent review.  REP thanks the
KITP in Santa Barbara for a stimulating environment enjoyed during the
composition of this chapter. RB is supported by the Deutsche
Forschungsgemeinschaft under grant KL 1358/4-1. REP and RO are
supported by grants from the National Science and Engineering Research
Council of Canada.


\input{Proceeding_Outflows_astroph.bbl}
\end{document}

%% file: journals.tex
\def\jnl@style#1{{\rmfamily#1}}%
\def\jref@jnl#1{{\jnl@style#1}}%

\newcommand\aj{\jref@jnl{AJ}}%
\newcommand\araa{\jref@jnl{ARA\&A}}%
\newcommand\apj{\jref@jnl{ApJ}}%
\newcommand\apjl{\jref@jnl{ApJ}}%
\newcommand\apjs{\jref@jnl{ApJS}}%
\newcommand\ao{\jref@jnl{Appl.~Opt.}}%
\newcommand\apss{\jref@jnl{Ap\&SS}}%
\newcommand\aap{\jref@jnl{A\&A}}%
\newcommand\aapr{\jref@jnl{A\&A~Rev.}}%
\newcommand\aaps{\jref@jnl{A\&AS}}%
\newcommand\azh{\jref@jnl{AZh}}%
\newcommand\baas{\jref@jnl{BAAS}}%
\newcommand\jrasc{\jref@jnl{JRASC}}%
\newcommand\memras{\jref@jnl{MmRAS}}%
\newcommand\mnras{\jref@jnl{MNRAS}}%
\newcommand\pra{\jref@jnl{Phys.~Rev.~A}}%
\newcommand\prb{\jref@jnl{Phys.~Rev.~B}}%
\newcommand\prc{\jref@jnl{Phys.~Rev.~C}}%
\newcommand\prd{\jref@jnl{Phys.~Rev.~D}}%
\newcommand\pre{\jref@jnl{Phys.~Rev.~E}}%
\newcommand\prl{\jref@jnl{Phys.~Rev.~Lett.}}%
\newcommand\pasp{\jref@jnl{PASP}}%
\newcommand\pasj{\jref@jnl{PASJ}}%
\newcommand\qjras{\jref@jnl{QJRAS}}%
\newcommand\skytel{\jref@jnl{S\&T}}%
\newcommand\solphys{\jref@jnl{Sol.~Phys.}}%
\newcommand\sovast{\jref@jnl{Soviet~Ast.}}%
\newcommand\ssr{\jref@jnl{Space~Sci.~Rev.}}%
\newcommand\zap{\jref@jnl{ZAp}}%
\newcommand\nat{\jref@jnl{Nature}}%
\newcommand\iaucirc{\jref@jnl{IAU~Circ.}}%
\newcommand\aplett{\jref@jnl{Astrophys.~Lett.}}%
\newcommand\apspr{\jref@jnl{Astrophys.~Space~Phys.~Res.}}%
\newcommand\bain{\jref@jnl{Bull.~Astron.~Inst.~Netherlands}}%
\newcommand\fcp{\jref@jnl{Fund.~Cosmic~Phys.}}%
\newcommand\gca{\jref@jnl{Geochim.~Cosmochim.~Acta}}%
\newcommand\grl{\jref@jnl{Geophys.~Res.~Lett.}}%
\newcommand\jcp{\jref@jnl{J.~Chem.~Phys.}}%
\newcommand\jgr{\jref@jnl{J.~Geophys.~Res.}}%
\newcommand\jqsrt{\jref@jnl{J.~Quant.~Spec.~Radiat.~Transf.}}%
\newcommand\memsai{\jref@jnl{Mem.~Soc.~Astron.~Italiana}}%
\newcommand\nphysa{\jref@jnl{Nucl.~Phys.~A}}%
\newcommand\physrep{\jref@jnl{Phys.~Rep.}}%
\newcommand\physscr{\jref@jnl{Phys.~Scr}}%
\newcommand\planss{\jref@jnl{Planet.~Space~Sci.}}%
\newcommand\procspie{\jref@jnl{Proc.~SPIE}}%